# Collective Intelligence in Citizen Science – A Study of Performers and Talkers

Ramine Tinati, Elena Simperl, Markus Luczak-Roesch, Max Van Kleek, Nigel Shadbolt, University of Southampton

1. INTRODUCTION

*Online citizen science* can be seen as a form of collective intelligence Lévy [1997] and Woolley et al. [2010] in which the wisdom of the crowd is applied to the Web to advance scientific knowledge [Prestopnik and Crowston 2012] Thus far, online citizen science projects [Bonney et al. 2009] have applied millions of volunteers to solving problems in a wide array of scientific domains, ranging from the classification of galaxies [Fortson et al. 2011] to the completion of protein folding networks [Khatib et al. 2011].

Central to many of these projects are online messaging or discussion facilities designed to allow volunteers to ask one another questions and advice. Such facilities have in many cases yielded substantial, dedicated self-sustaining online communities. In this paper, we examine participation in such communities; specifically, whether participation in online discussion influences task completion within and across 10 distinct projects of a shared citizen science platform, the Zooniverse[1]. Our study was conducted on a dataset from December 2009 to July 2013, in which $250,000$ users contributed over $50$ million tasks and $650,000$ discussion posts.

2. RELATED WORK

The hybrid nature of online citizen science, as part Web-based community and part crowdsourcing exercise, places at the intersection of several important research disciplines, including data analysis, social computing, and collective intelligence. In this section we will give a brief overview of key works in these areas, which have informed our analysis.

Online communities have been studied from a variety of perspectives; from that of the individual to the entire system [Arguello et al. 2006], as peer-production systems and as information sharing repositories [Krieger and Stark 2009; Slattery 2009], exhibiting emergent characteristics that develop and evolve with time [Kumar et al. 2006]. Closer to our research, which investigates the role of discussion forums in citizen science, is the literature around question-answering systems [Harper et al. 2009] and online collaboration platforms, most importantly Wikipedia [Kittur et al. 2007]. There is an extensive body of work that proposed models, methods, and case studies considering aspects such as an answer quality [Agichtein et al. 2008]; topics coverage and commonly used language [Rowe et al. 2013]; as well as user profiles, roles, and expertise [Fisher et al. 2006; Pal et al. 2012].

Independently of the underlying community, citizen science projects can be roughly characterized as being primarily concerns with data collection, or data processing and analysis [Kawrykow et al. 2012]. While their primary purpose is to solve large-scale data-intensive research problems, they also aim at creating a sustainable community of engaged amateur scientists that is willing and skilled to contribute to future projects. Studies of such projects have so far focused on human-computation aspects, describing specific mechanisms for task design, quality assurance, or coordination [Kawrykow et al. 2012; Khatib et al. 2011], as well as on participation, studying what drives the users to contribute [Raddick et al. 2010; Rotman et al. 2012].

---

[1] http://www.zooniverse.org

Compared to these lines of research, our work is different in two important ways: we use content and social network analysis methods to look at a class of online communities that so far remains underexplored, while offering the first quantitative, cross-project study of participation in citizen science.

## 3. THE ZOONIVERSE PLATFORM

'Zooniverse' is a citizen science Web platform which hosts, as of January $2014$, $30$ separate citizen science projects spanning several scientific and humanities domains. By using collective intelligence techniques, participants contribute to projects by performing data classification and analysis tasks on digital artifacts in image, video or audio form. An artifact is viewed by multiple participants and the results are validated using a combination of machine learning techniques and expert validation by the scientific team leading the project. Facilitating the participants, each project is linked with a discussion forum and messaging facilities called *Talk*, which serve as the main tool for information sharing and social interaction among participants.

## 4. RESULTS

First we summarise an analysis of the relationship between Talk participation and tasks performed across 10 Zooniverse projects. The analysis described here is a summary of early observations resulting from work[2] which provides a more thorough, multi-perspective analysis.

### 4.1 Talk versus Tasks: Do Those Who Discuss Contribute More?

We found that of the $250,071$ participants in our data, only 40.5% of them had contributed both classifications and discussions. Since some of the participants may have not been aware of the discussion facilities, we restricted our analysis to to those who posted at least once to avoid skewing our results. Figure 1 compares the number of classifications performed per user (x-axis) with the number of posts they made in the Talk system (y-axis). While the overall positive trend suggests that those who are generally more active perform more posts and tasks, participants completed vastly more tasks than made posts, performing a median 600 classifications versus 14 discussion posts each.

However, there exists a collection of participants that not only have performed a much larger number of classifications, but also have participated extensively in the discussion forums. To focus on such highly active participants, we extracted a subset that had performed at least $140$ posts and $6000$ classifications. This contained $2928$ users across all projects, which in total represented 29.0% and 72.0% of the total number classification and discussions in our data set, respectively. Within this set, we found three different pattern of participants, (1) individuals that sporadically perform classifications, (2) individuals that consistently perform classifications, (3) individuals that perform classifications very infrequently. Then, by using a k-means approach and clustering the participants by their frequency of Talk entries and task completion, we identified five characteristic user types illustrated in Figure I.

### 4.2 Roles in the Discussion Forums

A second analysis pertained to roles played by participants in Talk discussions. While superficially, much of the role of participants in forum discussions consisted of fielding and answering questions, such questions and the conversations around them often served different purposes. Figure II illustrates an initial set of roles we identified, which illustrates a considerable breadth in the kinds of activities performed. While these roles overlap considerably with roles identified in online QA communities such as [Pal et al. 2012], the emergent roles of *Discoverer*, *Hypothesiser*, *Investigator*, were distinctive to citizen science that were not only crucial to the collaborative hypothesis process, but demonstrated that the citizens themselves were engaging in a scientific process. This was supported by an extended analysis of the change in vocabulary used by participants, which indicated an increase in topic expertise within certain projects [Luczak-Roesch et al. 2014].

---

[2]Why Won't Aliens Talk to Us? Content and Community Dynamics in Online Citizen Science [Luczak-Roesch et al. 2014]

| User type | Description |
| --- | --- |
| Casual Hobbyists | Moderate task completion making occasional Talk entries sustained over a long time, with slow participation decline |
| Short but Sweet | Highly active in both task and Talk, but short-lived participation |
| Birth of a Moderator | More classifications at the beginning with increasing involvement in Talk activities over time |
| Performer to Talker | High number of tasks performed to begin with, then starts to use Talk instead |
| Talker to Performer | Frequent use of Talk before engaging with tasks |

Table I. : User behaviour stereotypes identified among the most active participants.

| Participant Role | Description |
| --- | --- |
| *General Help Asker* | Asked a general clarification or help question |
| *Answerer* | Answers a question with a definitive answer. |
| *Informer* | Posts unsolicited (but often authoritative) information to the community. |
| *Moderator* | Encourage/discourage certain posting behaviour, redirecting participants to a different thread or location where |
| *Discoverer* | Asked a question ultimately resulted in an unprecedented discovery |
| *Hypothesiser* | Poses a scientific question or proposes a possible answer. |
| *Investigator/Validator* | Took hypotheses posed and did further investigation/validation, presenting these results |
| *Cheerleader* | Encourages group to continue performing tasks, such as to reach milestones |
| *Celebrator* | Points to interesting objects for their aesthetic/fun value |

Table II. : Roles in discussion forums

## 5. CONCLUSIONS

In this paper, we examined the phenomenon of citizen science through the lens of the Zooniverse platform. The central contributions are the insights gained in understanding the relationship between Talk and task completion in respects to online citizen science, and the discovery if emergent user roles. We discovered a relationship between those that completed tasks and participated in Talk discussions and found a set of 'active' users which were responsible for over 70% of the Talk content, thus assuming the role of the 'core community'. These results reflect other peer-production systems like Wikipedia; despite obtaining a large user-base, it is the activities of only a relatively small collection of users that produce content.

In addition to this study, our ongoing research of online citizen science has found noticeable cross-project phenomenon; we are discovering how language of communities evolves relative to their subject domain, and how serendipitous discoveries are achieved via collective behaviour. Ultimately, this has wider implications for understanding collective intelligence on the Web. Given that the current landscape of crowdsourcing is somewhat of a disjoint collection of communities, our findings suggest that it is beneficial to support and unite online communities within and between different crowdsourcing systems, independent of subject domain or topic.

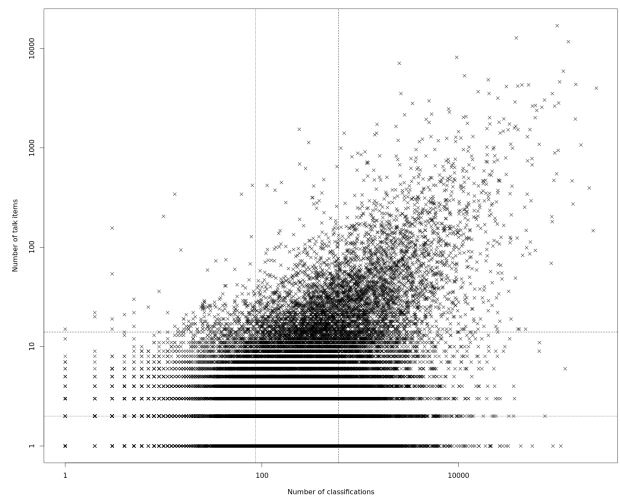

Fig. 1: Number of classifications (x-axis) and talk contributions (y-axis) of every single user.